\documentclass[
 superscriptaddress,
 reprint,
 amsmath, amssymb,
 aps,prl, preprintnumbers, nofootinbib, floatfix, longbibliography
]{revtex4-1}

\usepackage{graphicx}
\usepackage{dcolumn}
\usepackage{hyperref}
\usepackage{bm}
\usepackage{ulem}
\usepackage{color}
\usepackage{amssymb}
\usepackage{amsmath}
\usepackage{comment}
\usepackage{adjustbox}
\usepackage{rotating}
\usepackage{multirow}
\usepackage{graphicx}
\usepackage{lipsum}
\usepackage{slashed}
\usepackage{tabularray}
\begin{document}
\preprint{}
\title{Instantaneous Symmetry Breaking to the Non-observed Dark Matter}

\author{Y. Fang}
\email{fangyq@ihep.ac.cn}
\affiliation{Institute of High Energy Physics, Chinese Academy of Sciences, Beijing 100049, China}
\affiliation{University of Chinese Academy of Sciences, Beijing 100049, China}



\begin{abstract}
A theory explaining the non-observation of the dark matter and the source of the dark energy is presented in this letter. By integrating the asymmetrical potential and the Higgs potential, we provide a model with instantaneous symmetrical breaking and stable symmetrical breaking, resulting in the non-observed dark matter and observed matter respectively. Two crucial parameters in this model are the frequency and strength of the symmetry breaking from the vacuum: the former helps explain the impact of the effective mass from the dark matter; the latter determines the source of the dark energy. The expected strength in a certain period varies, causing the accelerating or deccelerating expansions of the universe. Considering the expected strength correlated with the vacuum expectation value and basing on the possible variations of the measured masses of the fundamental particles such as Z boson over time, one can perhaps derive the exact phase of the current universe.

\end{abstract}

\maketitle



\paragraph{Introduction}
With the discovery of the Higgs~\cite{ATLAS2012yve,CMS2012qbp} at the Large Hadron Collider (LHC), the mechanism that breaks electroweak symmetry in the Standard Model (SM)~\cite{PhysRevLett.13.321,HIGGS1964132,PhysRevLett.13.508,PhysRevLett.13.585,PhysRev.145.1156,PhysRev.155.1554} has been verified experimentally. While the SM has been remarkably successful in describing experimental phenomena, it is important to recognize that it is not a complete theory. A lot of issues or questions are still puzzles for particle physicists: Anything fundamental stays behind flavor symmetry? Mass hierarchy of elementary particles is normal? Dirac or Majorana Neutrino is massive?  Why cannot we observe dark matter while it is predicted according to the speed of the expansion of the universe? In this letter, we present a mechanism to explain the non-existence of the observed dark matter and the source of the dark energy.

\begin{figure}[h!]
	\centering
	{\includegraphics[width=.20\textwidth]{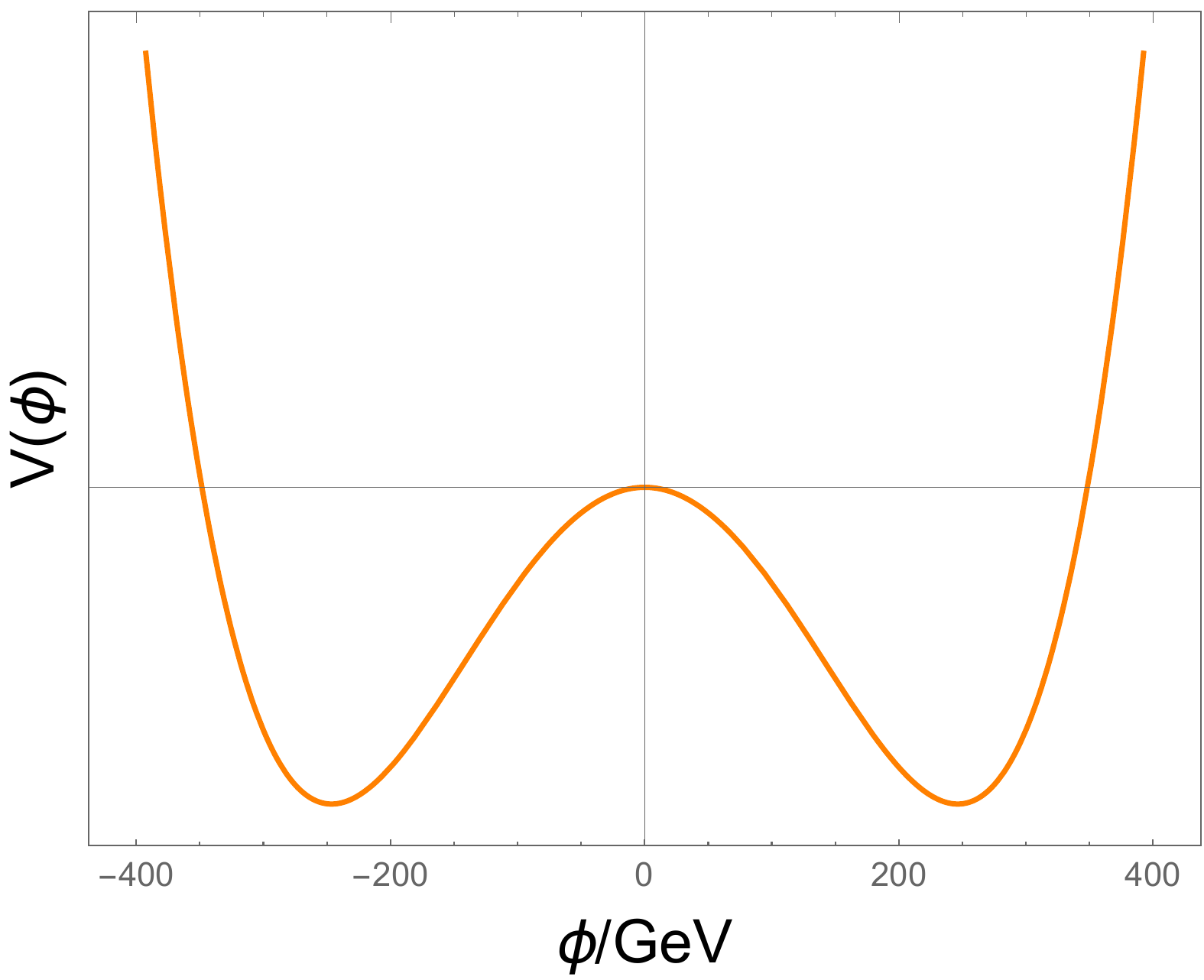}}
	{\includegraphics[width=.20\textwidth]{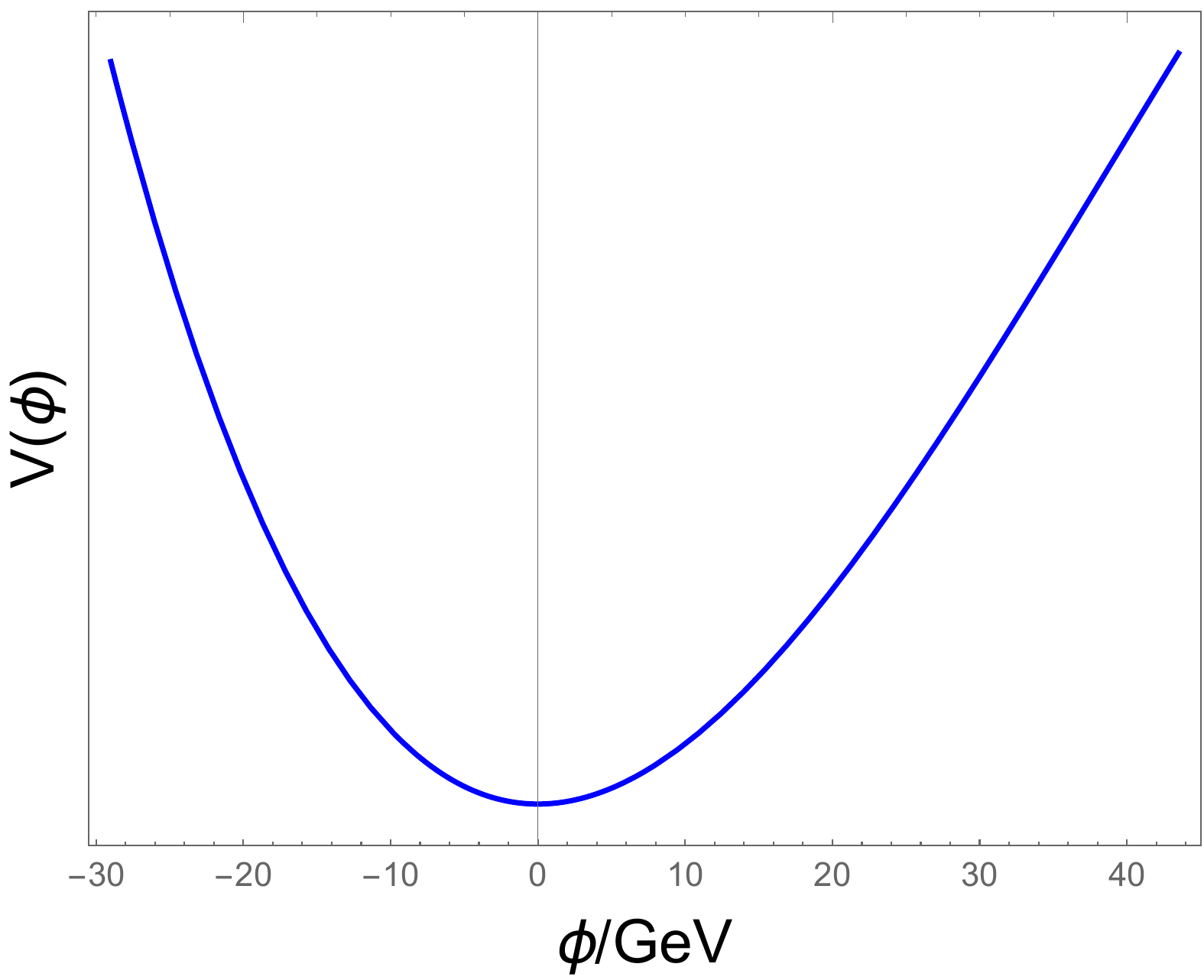}}
	\caption{Left plot shows a potential leading to a stable symmetry breaking to generate the detectable mass. Right plot shows a potential not symmetrical around Y-axis and causes an instantaneous symmetry breaking with a zero VEV. }
	\label{fig:potential1}
\end{figure}

Vacuum can be viewed as a complete symmetry. Mass can be generated from the vacuum with the fluctuations of spontaneous symmetry breaking.  It is obvious that a potential of Eq.~\ref{eq:potential1}, i.e. Higgs potential, will lead to a stable symmetry breaking (SSB) and hence  the generation of the detectable mass, i.e. a non-zero vacuum expectation value (VEV), as the left plot of Fig.~\ref{fig:potential1} describes. 

\begin{equation}
\label{eq:potential1}
V_{\phi}=-\frac{m_h^2}{4}\phi^{2} + \frac{m_h^2}{8v^2}\phi^{4}
\end{equation}

However, a potential asymmetrical around Y-axis, as shown in the right plot of Fig.~\ref{fig:potential1}, can cause an instantaneous symmetry breaking (ISB) but yielding a zero VEV.  This will therefore result in non-observed mass. 

\begin{figure}[h!]
	\centering
	{\includegraphics[width=.45\textwidth]{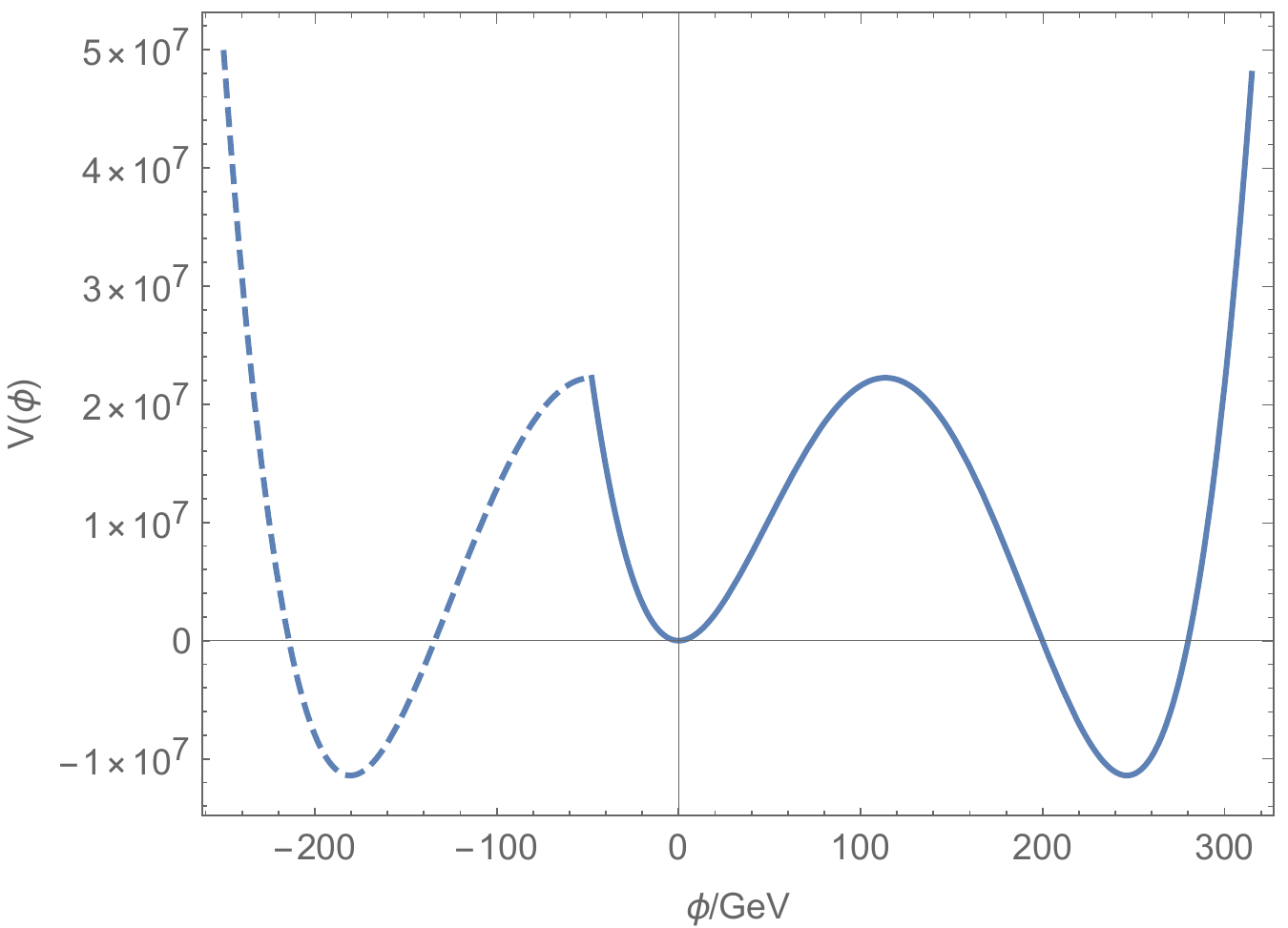}}
	\caption{Potential to model the production of the matter and dark matter. }
	\label{fig:potential2}
\end{figure}

The potential of the second scenario is unfortunately most likely non-renormalizable. A potential illustrated by the solid curve in Fig.~\ref{fig:potential2}, not only behaves renormalizablly, but also incorporates the properties of the two potentials shown in Fig.~\ref{fig:potential1} by plugging in the dashed curve which is the mirror image of the right partner. It can be potentially employed to model the productions of the matter and dark matter. 

For example, the potential proposed with the solid curve in Fig.~\ref{fig:potential2} can be written as
\begin{equation}
\label{eq:potential2}
V_{\phi}=\lambda\phi^{2}(\phi-\Lambda_1)(\phi-\Lambda_2)~,
\end{equation}
which satisfies
\begin{equation}
0<\Lambda_1<\Lambda_2~.
\end{equation}
The minima of $V(\phi)$ are located at
\begin{equation}
\left\langle\phi\right\rangle=0
\end{equation}
and 
\begin{equation}
\left\langle\phi\right\rangle=\frac{3}{8}(\Lambda_1+\Lambda_2+\Lambda^\prime)~,
\end{equation}
where
\begin{equation}
\Lambda^\prime=\sqrt{\Lambda_1^2+\Lambda_2^2-\frac{14}{9}\Lambda_1 \Lambda_2}~.
\end{equation}
To ensure the minimum of $V(\phi)$ with non-vanish $\left\langle\phi\right\rangle$ with the same $v$ and $m_h$ as those in SM, it is needed that
\begin{equation}
\left\{\begin{array}{l}
\Lambda_1=\frac{-m_h^2+8 \lambda v^2-\sqrt{m_h^4+2 \lambda m_h^2 v^2-8 \lambda^2 v^4}}{6 \lambda v} \\
\Lambda_2=\frac{-m_h^2+8 \lambda v^2+\sqrt{m_h^4+2 \lambda m_h^2 v^2-8 \lambda^2 v^4}}{6 \lambda v}
\end{array}\right.
\end{equation}
and   
\begin{equation}
\frac{m_h^2}{4 v^2}<\lambda<\frac{m_h^2}{2 v^2}~.
\end{equation}

The overall picture is as following; the vacuum with the potential shown in Fig.~\ref{fig:potential2} vibrates: the symmetrical breaking with higher strength but lower frequency ending up in one of the pitfalls with VEV larger than zero results in the massive particles. Those fluctuations with much higher frequency but lower strength within the asymmetrical center well lead to zero observed mass considering the VEV equal to zero. However, due to the asymmetrical property of the potential and possible tremendous high frequency of the ISB, it renders the effect of mass, namely non-observed dark matter, contributing in the gravitational force but not involving in the interactions with other particles.

If the central potential in Fig.~\ref{fig:potential2} is absorbed to one point, the model can be simplified as the Higgs potential as the left plot of Fig.~\ref{fig:potential1} displays.  The strength and frequency of the fluctuations are two parameters to describe the ISB/SSB of the vacuum. For observable matter, VEV can be used to describe the expected strength for the SSB; In the other word, the higher strength leads to the higher VEV, thus the heavier observable fundamental particles and vice versa. The central potential of Fig.~\ref{fig:potential2} can be simplified as a double-Gaussian and the average width for the double-Gaussians can be treated as its effective expected strength; while the difference of the widths is proportional to the effective mass of the dark particles.  
The expected strength multiplied by the frequency is defined as the capacity of the symmetry breaking of the vacuum. This capacity is assumed to be conserved, which means if the production of the strength and frequency for the matter decreases, the production for dark matter increases and vice versa.  

The expected strength in certain period of the ISB of the vacuum is the source of the dark energy and also plays the major role for the expansion of the universe. If taking the widely accepted fractions  of dark energy, dark matter and matter which are 68\%, 27\% and 5\%, one can have :
\begin{align}
\label{eq:delta1}
&\frac{f_{d}(W_{R}-W_{L})}{ {f_{d}(W_{R}+W_{L})/2}}= \frac{27\%}{68\%}\\
&\frac {f_{d}(W_{R}-W_{L})}{f_{m}V} =  \frac{27\%}{5\%}
\end{align}
where $W_{R}$ and $W_{L}$ are the right and left width of the central double-Gaussian potential and V represents the VEV approximately 246 GeV. $f_{d}$ and $f_m$ are the frequencies of the vibrations for the dark matter and symmetrical breaking for the matter. It is obvious that $W_{R}$ $\sim$ 1.5 $W_{L}$ and  
\begin{align}
\label{eq:delta22}
&\frac {W_{L}}{V} \sim  10.8 \frac{f_{m}}{f_{d}}
\end{align}

Taking into account $W_{L}/V$ has to be smaller than 1, $f_{d}$ is at least one order of magnitude higher than $f_{m}$. With the reasonable assumption that $W_{L}$ is around or lower than O(1) MeV, the order of the mass of the electron within the SSB family,  the ratio between $f_{d}$ and $f_{m}$ is amazingly at least O($10^{6})$! It confirms the assumption that the frequency of the ISB is much larger than SSB.    

The vibration of the above strength leads to the expansion as well as the contraction of the universe 
which rotates with four phases: expanding acceleratively, expanding deceleratively, contracting acceleratively, contracting deceleratively. In this paper, only the expansion of the universe is discussed.  The strength of the spontaneous breaking of the vacuum is supposed to vary, indicating the variation of the VEV and thus the mass deviations of the heavy fundamental particles measured at the colliders, e.g. W, Z, Higgs and top quark etc., from the former measurements to the latter ones with the formula: 
\begin{align}
\label{eq:delta1}
&{m} = \alpha {V}\\
&\frac {\Delta m}{m} =  \frac{\Delta V}{V}
\end{align}
where $m$ is the mass of the fundamental particle. $\Delta m$ and $\Delta V$ are the deviation of $m$ and $V$ due to the above strength variation. $\alpha$ are the coupling coefficients, whose approximate values for different fundamental particles are listed in Table \ref{table1}. Assuming $\Delta V/V\sim 10^{-4}$, the same level of the average degradation of $W$ boson mass from Tevetron (D0/CDF)~\cite{cdfWmass,D0:2012kms} to the LHC(ATLAS/LHCb)~\cite{atlasWmass,LHCbbjt} later,  the expected deviations on $m$ and current uncertainties are listed in the third and fourth columns respectively in Table \ref{table1}.

\begin{table}
\centering
\begin{tblr}{
  cells = {c},
  cell{1}{1} = {r=2}{},
  cell{1}{2} = {r=2}{},
  cell{1}{3} = {c=2}{},
  vlines,
  hline{1,3-9} = {-}{},
  hline{2} = {3-4}{},
}
Particle & $\alpha$ & $\Delta m$/GeV   \\
 &  & deviation & current uncertainty \cite{Workman:2022ynf}\\
$W$ & 0.327 & $8.04\times 10^{-3}$ & $1.2\times 10^{-2}$\\
$Z$ & 0.371 & $9.12\times 10^{-3}$ & $2.1\times 10^{-3}$\\
$H$ & 0.509 & $1.25\times 10^{-2}$ & 0.17 \\
top & 0.702 & $1.73\times 10^{-2}$ & 0.30
\end{tblr}
	\caption{The expected deviations of the masses for the fundamental particles assuming the variation of the VEV is $\Delta V/V\sim 10^{-4}$.}
	\label{table1}
\end{table}

In practice, the high precision measurements of Z mass around 2 MeV at the Large Electron Positron Collider (LEP)~\cite{lepzmass} in 1990s and the LHC running now could provide some evidences depending on the achieved systematic uncertainties and the deviations.  

Of course, the future electron positron colliders such as the Circular Electron Positron Collider (CEPC)~\cite{CEPCStudyGroup:2018ghi,CEPCPhysicsStudyGroup:2022uwl} and the Future Circular Lepton Collider (Fcc-ee)~\cite{FCC:2018byv} are idea facilities to deliver the Z mass measurements with an accuracy less than 1 MeV, better than that from the LEP. Based on these, the exact phase of the current universe could be identified. This means an accelerating expansion of the universe at the present time if Z mass increases; Otherwise, a decelerating expansion of the universe expected. The latter scenario is consistent with the trend of current measurements of the top mass from  $174.34\pm0.64$ GeV (Tevetron)~\cite{CDF:2014upy} to $172.52\pm0.33$ GeV (LHC)~\cite{topmassLHC} besides the measured $W$ mass reduction mentioned above. Furthermore, the measurements of Rydberg Constant from 2014 to 2018 ~\cite{massele} tend to reduce with a significance of $5.3\sigma$, indicating a reduced mass of electron over time. 

As long as the universe expands with lower and lower acceleration, the effective dark matter is supposed to increase due to less energetic symmetrical breaking of the vacuum going beyond the central potential. Therefore the stability of the universe is still likely to be kept.

\begin{acknowledgments}

The authors thank to the discussions and encouragements by J. Pei, X. Lou, Y. Jiang, Y. Guo and P. Huang. The study is supported partially by Institute of High Energy Physics, Chinese Academy of Sciences under the innovative project on sciences and technologies with no. E3545BU210. 
\end{acknowledgments}

\bibliography{sample}


\end{document}